\documentclass[useAMS,usenatbib]{mn2e}

\usepackage{lscape}
\usepackage{supertabular}
\usepackage{longtable}
\usepackage{graphicx}
\usepackage{txfonts}
\def\gtsima{$\; \buildrel > \over \sim \;$}
\def\ltsima{$\; \buildrel < \over \sim \;$}
\def\gsim{\lower.5ex\hbox{\gtsima}}
\def\lsim{\lower.5ex\hbox{\ltsima}}

\title[Relation between black hole mass and soft X-ray lags in AGN]     
  {Discovery of a relation between black hole mass and soft X-ray time lags in active galactic nuclei}          
\author[B. De Marco et al]
  {B.~De Marco,$^{1,2,7}$\thanks{E-mail: demarco@iasfbo.inaf.it} 
  G.~Ponti,$^3$ M.~Cappi,$^2$
  M.~Dadina,$^2$  P.~Uttley,$^4$, E. M.~Cackett$^{5,6}$,
  \newauthor A. C.~Fabian,$^6$ and G.~Miniutti$^7$\\ 
  $^1$Dipartimento di Astronomia, Universit\`a di Bologna, via Ranzani 1, I-40127 Bologna, Italy\\
  $^2$Istituto di Astroﬁsica Spaziale e Fisica Cosmica-Bologna, INAF, via Gobetti 101, I-40129 Bologna, Italy\\
  $^3$Faculty of Physical and Applied Science, University of Southampton, Southampton SO17 1BJ\\
  $^4$Astronomical Institute ``Anton Pannekoek'', University of Amsterdam, Postbus 94249, 1090 GE Amsterdam, The Netherlands\\
  $^5$Wayne State University, Department of Physics and Astronomy, Detroit, MI, 48201, USA\\
  $^6$Institute of Astronomy, Madingley Road, Cambridge CB3 0HA\\
  $^7$Centro de Astrobiolog\'{i}a (CSIC-INTA), Dep. de Astrof\'{i}sica; ESAC, PO Box 78, E-28691, Villanueva de la Ca\~nada, Madrid, Spain}
\date{Released 2011 Xxxxx XX}

\pagerange{\pageref{firstpage}--\pageref{lastpage}} \pubyear{2011}

\def\LaTeX{L\kern-.36em\raise.3ex\hbox{a}\kern-.15em
    T\kern-.1667em\lower.7ex\hbox{E}\kern-.125emX}

\begin{document}

\label{firstpage}

\maketitle

\begin{abstract}
  We carried out a systematic analysis of time lags between X-ray energy bands in a large sample (32 sources) of unabsorbed, radio quiet active galactic nuclei (AGN), observed by XMM-{\it Newton}. The analysis of X-ray lags (up to the highest/shortest frequencies/time-scales), is performed in the Fourier-frequency domain, between energy bands where the soft excess (soft band)
and the primary power law (hard band) dominate the emission.
We report a total of 15 out of 32 sources displaying a high frequency soft lag in their light curves. All 15 are at a significance level exceeding 97 per cent and 11 are at a level exceeding 99 per cent. Of these soft lags, 7 have not been previously reported in the literature, thus this work significantly increases the number of known sources with a soft/negative lag.
The characteristic time-scales of the soft/negative lag are relatively short (with typical frequencies and amplitudes of $\nu\sim 0.07-4 \times 10^{-3}$ Hz and $\tau\sim 10-600$ s, respectively), and show a highly significant ($\gsim 4\sigma$) correlation
with the black hole mass. The measured correlations indicate that soft lags
are systematically shifted to lower frequencies and higher absolute amplitudes as the mass of the source increases. 
To first approximation, all the sources in the sample are consistent with having similar mass-scaled lag properties.
These results strongly suggest the existence of a mass-scaling law for the soft/negative lag, that holds for AGN spanning a large range of masses 
(about 2.5 orders of magnitude), thus supporting the idea that soft lags originate in the innermost regions of AGN and are powerful tools for testing their physics and geometry.

\end{abstract}


\section{Introduction}
\label{intro}

The observed similarities in the timing properties of different black hole (BH) systems suggest that the same physical mechanism is at work in sources spanning a wide range of masses (e.g. Uttley et al 2005, McHardy et al 2006).
This observational fact represents an important achievement in the context of the theory of unification of BH accretion. 
 A breakthrough in this respect was the discovery of the mass-scaling law that regulates the characteristic time-scales of variability, 
identified with the high frequency break in the power spectral density function (PSD). This relation holds for objects of widely different size
 (i.e. over about eight order of magnitudes in mass, from BH X-ray binaries, BHXB, up to active galactic nuclei, AGN, McHardy et al 2006, Koerding et al 2007)
and is in agreement with expectations from standard accretion disk models, whereby all the characteristic time-scales depend linearly on the BH mass, $\mathit{M}_{\mathrm{BH}}$ (e.g. Shakura \& Sunyaev 1973, Treves et al 1988).\\
 Another fundamental analogy emerges from the comparison of time lags between X-ray energy bands. Hard/positive lags (i.e. hard X-ray  variations lagging soft X-ray variations) are generally observed in both BHXBs and AGN, and can be interpreted in terms of propagation of mass accretion rate 
fluctuations in the disc (Kotov et al 2001 and Ar\'{e}valo \& Uttley 2006).
Those lags are usually detected at relatively low fequencies 
(i.e. below the PSD high frequency break).\\
 On the other hand, a relatively new and interesting perspective comes from the study of high frequency, soft/negative lags (i.e. 
soft X-ray band variations lagging hard X-ray band variations). 
The first significant detection of a soft/negative lag in AGN light curves was reported in the narrow line Seyfert 1 (NLSy1) galaxy
1H0707-495 (Fabian et al 2009, Zoghbi et al 2010), and interpreted as the signature of relativistic reflection which ``reverberates'' in response to continuum
changes after a time equal to the light-crossing time from the source to the reflecting region.
A different interpretation has been proposed in terms of a complex system of scatterers/absorbers located close to the line of sight, but at hundreds of gravitational radii, $\mathit{r}_g$, from the central source (Miller et al 2010), thus requiring us to have a special line of sight to the source.\\
Subsequently, soft lags have been observed in several other sources (e.g. Tripathi et al 2011, Emmanoulopoulos et al 2011, Zoghbi \& Fabian 2011). 
One of the largest-mass sources ($\mathit{M}_{\mathrm{BH}}\sim10^{7-8} \mathit{M}_{\odot}$) showing a soft/negative lag is PG 1211+143 (De Marco et al 2011),
whose lag spectrum appears ``shifted'' by about one order of magnitude (with the lag frequency, $\nu$, and amplitude, $\tau$, found at $\sim 1-6\times 10^{-4}$Hz and $\sim 10^{2}$ s, respectively)
with respect to those measured in other low mass sources (i.e. $\nu\sim 10^{-3}$Hz and $\tau\sim 20-30$ s, Emmanoulopoulos et al 2011). 
In De Marco et al (2011) we speculated that this difference could be simply explained if the time-scales in PG 1211+143 are scaled-up 
by the BH mass, estimated to be about 10-100 times the mass of the other AGN for which X-ray lags have been recorded 
so far. In this paper we explore this hypothesis, by studying the properties of the lag frequency spectrum 
in a large sample of sources, spanning a wide range of masses (about 2.5 decades). 
The time-scales of detected soft/negative lags are then used to study correlations with relevant parameters, such as luminosity and BH mass.

\section{The sample}

The sources analysed are extracted from the \emph{CAIXAvar} sample -- a subsample of the CAIXA sample by Bianchi et al (2009a) -- presented in Ponti et al (2012), which includes all the well-exposed, X-ray unobscured ($\mathit{N}_H<2\times10^{22}$cm$^{-2}$), radio quiet AGN observed by XMM-{\it Newton} in targeted observations as of June 2010. We considered all the sources having at least one observation with a longer than 40 ks exposure (41 sources), and selected those with published black hole mass, $\mathit{M}_{\mathrm{BH}}$, estimates (39 sources). Finally, we selected all the sources showing significant variability in their light curves. To this aim we made use of the excess variance estimates provided by Ponti et al (2012), computed by sampling the 2-10 keV light curve of each source in 40 ks segments (note that also single-interval measurements were considered), and assuming a time resolution of 250 s. All the sources having excess variance different from zero at $\gsim$2$\sigma$ confidence level were included in our sample, i.e. excluding all the sources consistent with having constant flux on the time-scales of interest for this work ($\lsim 40$ks).
 With the latter selection our final sample was reduced to a total of 32 sources. 
For each source all the available observations in the XMM-{\it Newton} archive have been used in our analysis, 
apart from those highly corrupted by background flares.
In the computation of the lag-frequency spectra, multiple observations have been combined to obtain better statistics.\\
Only data from the EPIC-pn camera were used\footnote{We checked whether the EPIC MOS data yield consistent results, finding a good agreement, although the quality of the lag spectra is significantly affected by the lowest statistics.}, because of its high effective area and S/N over the 0.3-10 keV energy band.
Data reduction was performed using XMM Science Analysis System (SAS v. 10.0), starting from the Observation Data Files (ODF)
 and following standard procedures.
Filtered events are characterised by PATTERN $\leq$4, and are free from background flares. 
Typical source extraction regions are 45 arcsec radius circles. Spectra were extracted and used to
 select the energy bands for the computation of time lags.
They were grouped to a minimum of 20 counts per bin, while response matrices were generated 
through the RMFGEN and ARFGEN SAS tasks. 
The analysis of the time series was carried out using routines implemented through IDL v. 6.1. Correction of sporadic count rate drops (usually involving single time bins) in the time series, which occur as a consequence of event losses due to telemetry dropouts, was performed by rescaling the count rate within the bin for the effective fractional time bin length.

\section{Lag vs frequency spectra}

\subsection{Analysis}
We computed time lag-frequency spectra between light curves in the soft and hard X-ray energy bands. The soft and hard energy bands were selected so as to single out energies dominated by the soft excess and the primary power law. To this aim we adopted a phenomenological approach, although we note that several tests have been done by slightly varying the selected energy bounds, and the obtained results are all consistent with each other.
Specifically, we first fitted the data in the 1-4 keV energy band with a simple power law absorbed by a cold column of gas (which accounts for any warm absorber-induced extra curvature at high energies), with all the parameters left free to vary. For the soft band, we fixed at 0.3 keV the low energy bound, and selected, as high energy bound, the energy at which the soft excess significantly deviates (at more than $3\sigma$) from the extrapolation of the best fit absorbed-power law. As the hard energy band we used the 1-5 keV range, excluding energies where significant deviations from the best fit power law were observed, e.g. due to the presence of troughs produced by complex absorption (at $\sim 1$ keV) or to the presence of a broad Fe K line red tail (at $\sim 4-5$ keV). Data above 5 keV were, in general, excluded to avoid contamination from components of the Fe K line complex.
A detailed compilation of all the adopted energy bands can be found in Tables \ref{tab:obs} and \ref{tab:obs2}.\\ 
The time lag-frequency spectra are computed following the techniques described in detail by Nowak et al (1999) and applied to XMM-{\it Newton} data in De Marco et al (2011). The time lag is derived from the formula $\tau(\nu)=\phi(\nu)/2 \pi \nu$, where $\phi (\nu)$ is the frequency-dependent phase of the average cross power spectrum between the soft and hard time series. The resulting lag-frequency spectra are rebinned multiplicatively, i.e. the bin size is set equal to $n-$times the frequency value, with step size, $n$, chosen between $\sim1.2-2$ depending on the quality of the data. Sources for which a negative lag was reported in the literature have been re-analyzed with our procedures.

\begin{figure*}
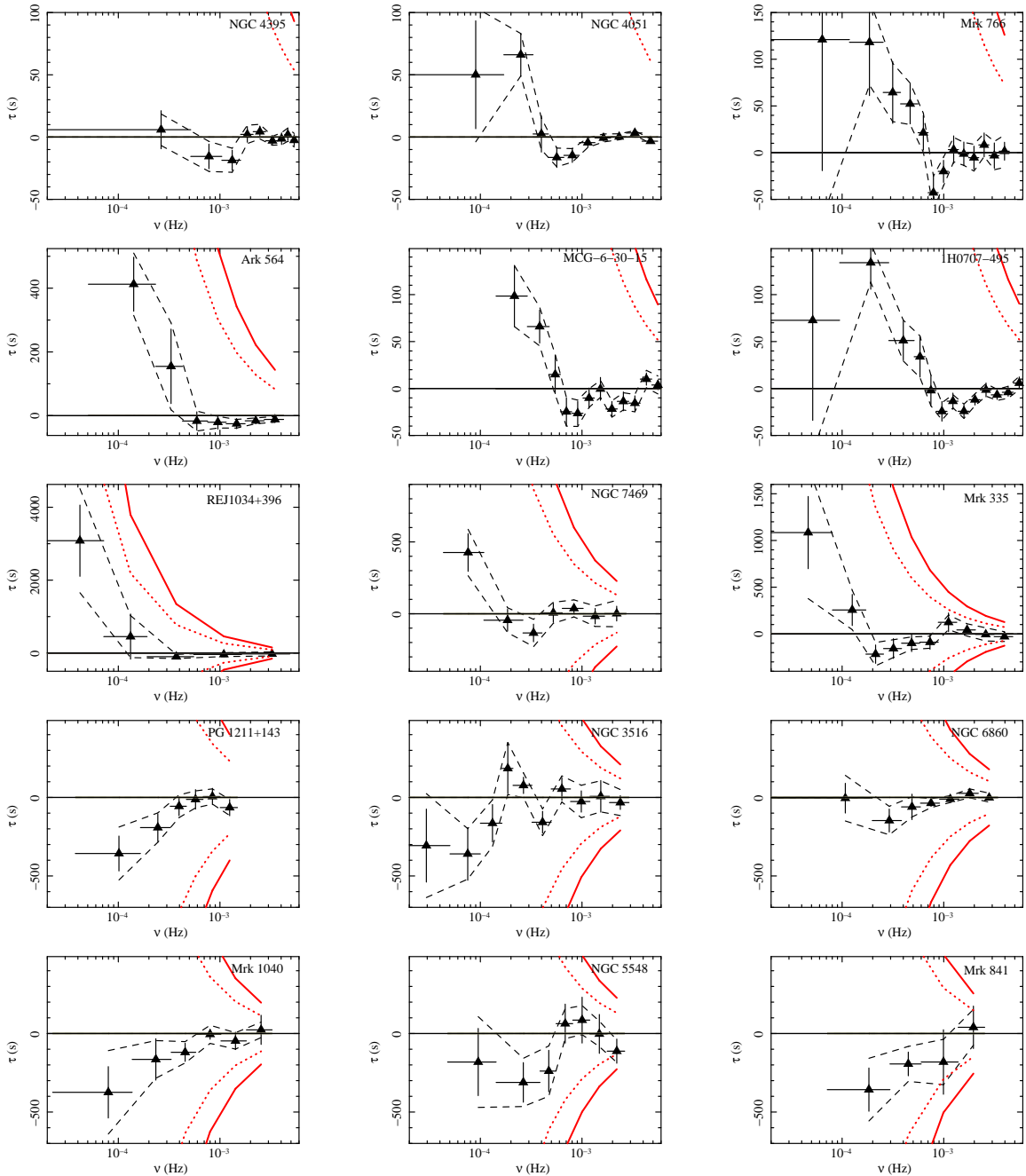

\centering
\begin{tabular}{p{5cm}p{5cm}p{5cm}}

\includegraphics[height=2.6cm,angle=270]{figure/NGC4395_lim2.ps} & \includegraphics[height=2.6cm,angle=270]{figure/NGC4051_lim2.ps} & \includegraphics[height=2.6cm,angle=270]{figure/Mrk766_lim2.ps}\\
 \includegraphics[height=2.6cm,angle=270]{figure/Ark564_lim2.ps} & \includegraphics[height=2.6cm,angle=270]{figure/MCG-6-30-15_lim2.ps} & \includegraphics[height=2.6cm,angle=270]{figure/1H0707_lim2.ps}\\
 \includegraphics[height=2.6cm,angle=270]{figure/REJ1034_lim2.ps} & \includegraphics[height=2.6cm,angle=270]{figure/NGC7469_lim2.ps} & \includegraphics[height=2.6cm,angle=270]{figure/Mrk335_lim2.ps}\\
 \includegraphics[height=2.6cm,angle=270]{figure/PG1211_lim2.ps} & \includegraphics[height=2.6cm,angle=270]{figure/NGC3516_lim2.ps} & \includegraphics[height=2.6cm,angle=270]{figure/NGC6860_lim2.ps}\\
\includegraphics[height=2.6cm,angle=270]{figure/Mrk1040_lim2.ps} & \includegraphics[height=2.6cm,angle=270]{figure/NGC5548_lim2.ps} & \includegraphics[height=2.6cm,angle=270]{figure/Mrk841_lim2.ps} \\
\end{tabular}
\caption{Lag-frequency spectra of sources with a detected soft/negative lag at $>2\sigma$ confidence level, in order of increasing mass (from top left to bottom right). The error bars are calculated using the standard formula from Bendat \& Piersol (1986), as used in Nowak et al (1999). The simulated $1\sigma$ contour plots (dashed curves) are overplotted on the data. The red-continous curves limit the range $\tau=[-1/2\nu,+1/2\nu]$ of allowed time lag values at each frequency, while the red-dotted curves mark the standard deviation of a uniform distribution defined on the same interval of time lag permitted values (see Sect. \ref{sec_pois} for details).}
\label{fig:lags}

\end{figure*}

\begin{figure*}
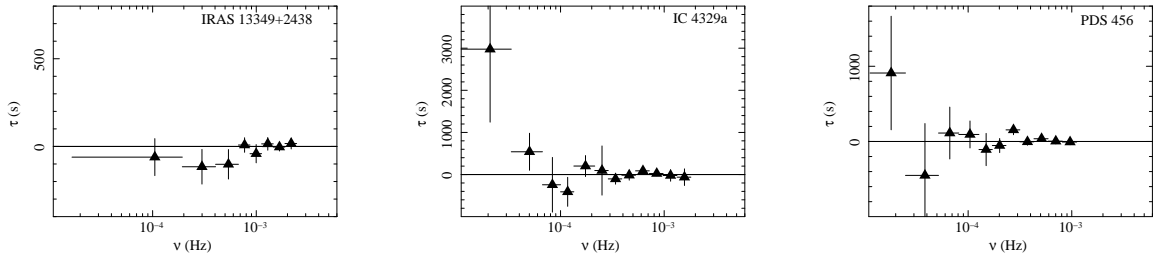
 
\centering
\begin{tabular}{p{5cm}p{5cm}p{5cm}}

\includegraphics[height=2.6cm,angle=270]{figure/IRAS13349.ps} & \includegraphics[height=2.6cm,angle=270]{figure/IC4329a.ps} & \includegraphics[height=2.6cm,angle=270]{figure/PDS456.ps} \\
\end{tabular}
\caption{Examples of lag-frequency spectra of sources with a marginal negative lag at $1.8\sigma$, $1.2\sigma$, and $<1\sigma$ (from left to right).}
\label{fig:lags2}

\end{figure*}

\begin{figure*}
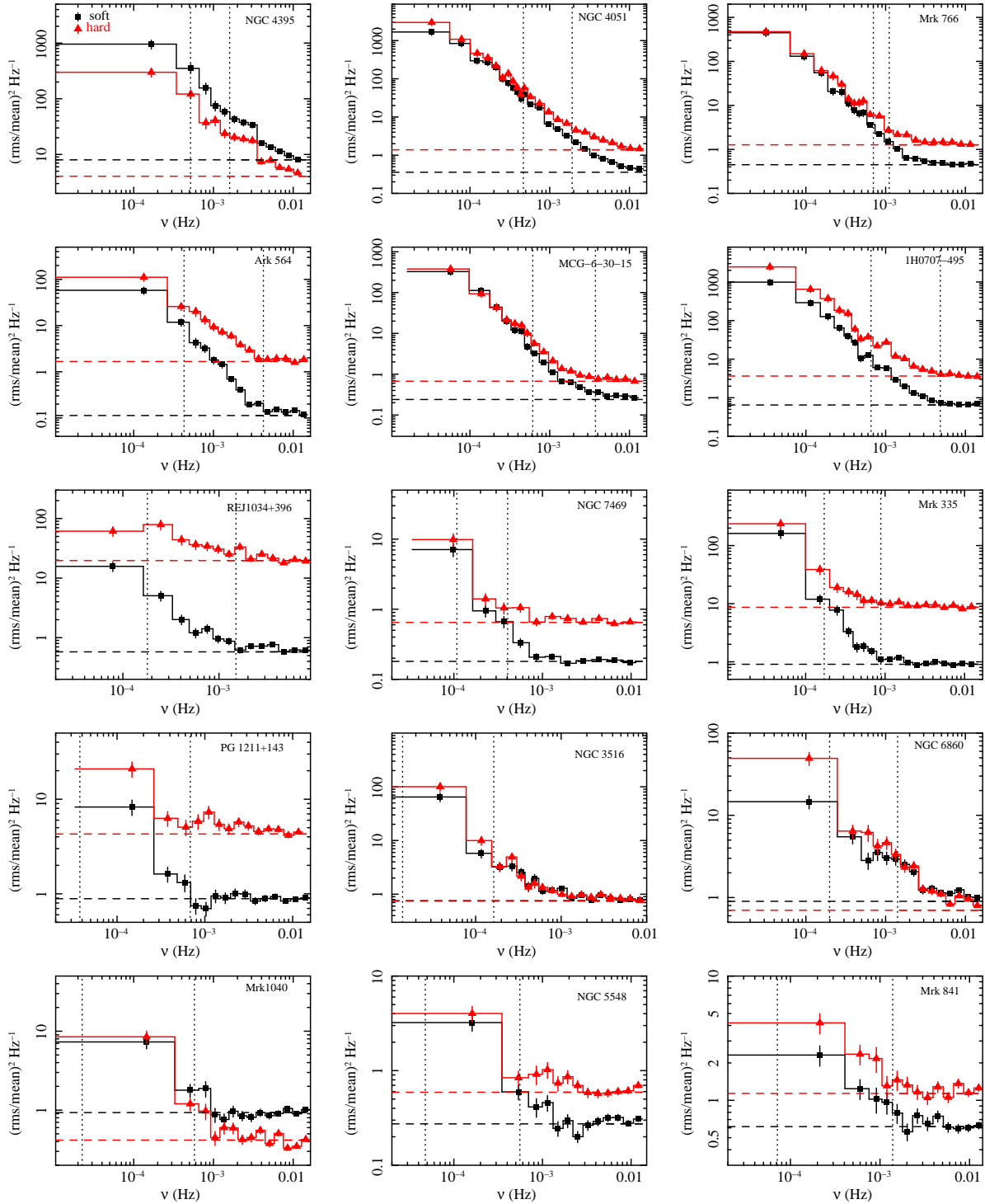
 
\centering
\begin{tabular}{p{5cm}p{5cm}p{5cm}}

\includegraphics[height=3.2cm,angle=270]{figure/NGC4395p_n.ps} & \includegraphics[height=3.2cm,angle=270]{figure/NGC4051p_n.ps} & \includegraphics[height=3.2cm,angle=270]{figure/Mrk766p_n.ps}\\

 \includegraphics[height=3.2cm,angle=270]{figure/Ark564p_n.ps} & \includegraphics[height=3.2cm,angle=270]{figure/MCG-6-30-15p_n.ps} & \includegraphics[height=3.2cm,angle=270]{figure/1H0707p_n.ps}\\

 \includegraphics[height=3.2cm,angle=270]{figure/REJ1034p_n.ps} & \includegraphics[height=3.2cm,angle=270]{figure/NGC7469p_n.ps} & \includegraphics[height=3.2cm,angle=270]{figure/Mrk335p_n.ps}\\

 \includegraphics[height=3.2cm,angle=270]{figure/PG1211p_n.ps} & \includegraphics[height=3.2cm,angle=270]{figure/NGC3516p_n.ps} & \includegraphics[height=3.2cm,angle=270]{figure/NGC6860p_n.ps}\\

\includegraphics[height=3.2cm,angle=270]{figure/Mrk1040p_n.ps} & \includegraphics[height=3.2cm,angle=270]{figure/NGC5548p_n.ps} & \includegraphics[height=3.2cm,angle=270]{figure/Mrk841p_n.ps} \\
\end{tabular}
\caption{Estimated power spectra (black squares: soft energy band; red triangles: hard energy band) of sources with a soft lag detected at $>2\sigma$ confidence level, in order of increasing mass (from top left to bottom right). The PSD are logarithmically rebinned (Papadakis \& Lawrence, 1993). The normalization defined by Miyamoto et al (1991) have been adopted. The vertical dotted lines mark the range of frequencies spanned by the soft lag, while the dashed horizontal lines mark the Poisson noise level for each band (computed following Vaughan et al 2003).}
\label{fig:psd}

\end{figure*}

\subsection{Results}
\label{sec:lag_specs}

We first present the results of our analysis, deferring to the next section a detailed discussion of their significance and robustness. We set as the detection threshold the 2$\sigma$ confidence level ($\sim$ 95 per cent) producing 15 out of 32 sources sources showing a soft lag with significance well above this limit (11 of them have significance $>99$ per cent, while the remaining 4 are above the $97$ per cent confidence level). The procedures followed to estimate the signficance of negative lag detections are discussed in detail in Sect. \ref{sec:lag_sign}.

The lag-frequency spectra of the 15 sources with a significant soft lag detection are shown in Fig. \ref{fig:lags}, while Fig. \ref{fig:lags2} displays some examples of spectra with only a marginal (significance $<2\sigma$) soft/negative lag. Most of the sources do show smooth lag profiles, several of which are characterised by a hard/positive lag at the lowest frequencies, dropping to negative values at higher frequencies, others showing only the soft/negative lag component, but shifted to low frequencies.\\
Our results are summarized in Tables \ref{tab:obs} and \ref{tab:obs2}, respectively for sources with a significant or marginal/non-significant detection. The detection significances reported in the Table \ref{tab:obs} are derived following different approaches (in particular, those enclosed in parentheses are obtained from extensive Monte Carlo simulations), as detailed in Sect. \ref{sec:lag_sign}.\\
Overplotted on the measured lag-frequency spectra (Fig. \ref{fig:lags}) are the 1$\sigma$ confidence levels of the observed time lags. The contours have been computed through Monte Carlo simulations (see Sect.\ref{sec:lag_sign}), using as a template-lag profile an interpolated function that describes the measured lag, and applying a frequency-dependent phase-shift to each pair of simulated light curves, according to this function. The spread in lag values is consistent with the errors associated with the data, which have been computed through the formula from Bendat \& Piersol (1986), as used in Nowak et al (1999).
Moreover, it is worth noting that, to recover the observed lag amplitude, we had to correct the input lag function by a frequency-dependent factor. If this correction is omitted, the lag-amplitudes at the lowest frequencies tend to be underestimated. This bias had already been pointed out in Zoghbi et al (2010) and De Marco et al (2011). It is related to a red noise leak effect, whereby phase-lag components associated with frequencies below the lowest sampled frequency contribute to those within the monitored frequency-window. If the lag associated with these components is either positive, null or smaller (in absolute value) than the intrinsic negative lag at the observable frequencies, the overall effect is of diluting the amplitude of the latter, thus underestimating a measurement. The amount of dilution depends on the amount of power below the lowest sampled frequency, thus being stronger for sources with steeper power spectra (e.g. the highest mass sources, for which the sampled frequency range in this analysis is far below the PSD high frequency break).\\
The soft/negative lags detected in our analysis are perfectly consistent (in both amplitude and frequency) with those already reported in the literature for eight of the sources of our sample (Ark 564, Ar\'{e}valo et al 2006; 1H0707-495, Zoghbi et al 2010; Mrk 1040, Tripathi et al 2011; Mrk 766, MCG-6-30-15, Emmanoulopoulos et al 2011; RE J1034+396, Zoghbi \& Fabian 2011; NGC 3516, Turner et al 2011; PG 1211+143, De Marco et al 2011), even when slightly different energy bands are adopted.
Moreover, we report a total of 7 newly detected soft lags (i.e. NGC 4395, NGC 4051, NGC 7469, Mrk 335, NGC 6860, NGC 5548, Mrk 841).
The observed values of lag amplitudes and frequencies span about two decades (i.e. $\tau\sim 10-600$ s and $\nu\sim 0.07-4 \times 10^{-3}$ Hz), similar to the sampled BH mass range of values ($\sim 0.1-10\times 10^7\mathit{M}_{\odot}$). These new detections significantly increase the number of soft/negative lags observed in nearby AGN.\\ 
Despite the wide range of BH masses and hence variability time-scales (and correspondingly, amplitudes) in our sample, it is interesting to note that we are still able to detect significant lags over a wide range of masses. One might expect that for the more slowly variable sources there would not be sufficient variable signal to detect the lags.  We address this question by showing the power spectral density function (PSD) of the 15 sources with a detected soft lag (see Fig. \ref{fig:psd}). The PSDs have been computed in the range of energies used to compute the lag-frequency spectra (i.e. in the soft and hard band selected for each source) so as to estimate the relative importance of variability power due to Poisson noise fluctuations with respect to the intrinsic power of the source at the lag frequencies. The lags are always detected at the frequencies where the Poisson noise contribution is negligible.  Thus it is not surprising that we are able to detect significant negative lags in all these sources. We refer to Sect. \ref{sec_pois} for further discussion about this issue.

\subsection{Robustness of lag detections}
\label{app}
\subsubsection{Negative lag significance}
\label{sec:lag_sign}

To determine the detection significance of the observed soft lags we used different methods. We first combined the significance of all the consecutive points that form the soft lag profile (using the error propagation rule for statistically uncorrelated data points) and derived the number of standard deviations from zero-lag. In the computation we did not take into account the low frequency hard/positive lag points (where present), and excluded frequencies where the effects of Poisson noise are significant (see details in Sect. \ref{sec_pois}). Indeed, the effect of counting noise is to produce a drop of coherence (the measure of the degree of linear correlation between the two light curves, e.g. Vaughan \& Nowak 1997) to zero-values. The corresponding cross spectrum phase at these frequencies is randomly distributed between $[-\pi,+\pi]$, thus precluding the detection of any intrinsic time lag. The soft/negative lags reported in this paper occur at frequencies well below the limit-frequency set by Poisson noise (Sect. \ref{sec_pois}), and are characterised by a medium-to-high level of intrinsic coherence ($0.4\lsim \gamma_{I}^2 \lsim 1$, estimated using prescriptions given by Vaughan \& Nowak 1997 for correction of counting noise effects), this value representing the fraction of the signal that is responsible for the lag. The derived soft lag significances have been cross-checked with those obtained by carrying out a $\chi^2$ test against a constant zero-lag model.\\
Noteworthy is the fact that, under a more rigorous approach, the single points in the lag-frequency profile cannot be treated as independent variables in standard statistical tests. For this reason the reliability of the soft lags in the 15 sources has been further tested through extensive Monte Carlo simulations. 
This approach allows us to check the robustness of the detections against spurious fluctuations produced by Poisson noise and red noise, over the entire sampled range of frequencies.
\begin{figure*}
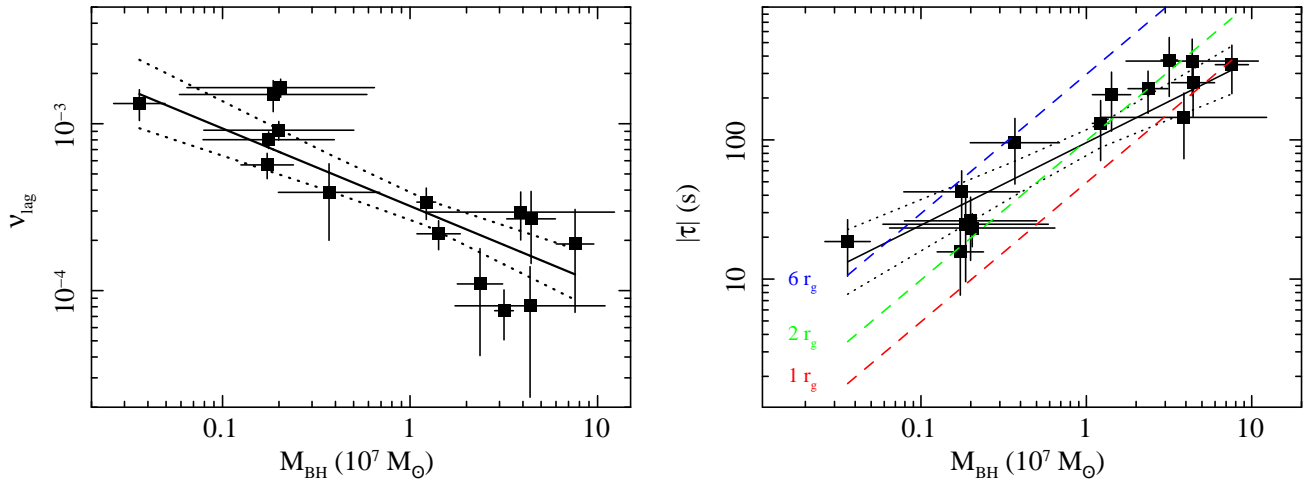
 
\centering

\begin{tabular}{p{8.5cm}p{8.5cm}}
\includegraphics[height=6.2cm,angle=270]{figure/corr_Mfreq_cont.ps} & \includegraphics[height=6.2cm,angle=270]{figure/corr_Mtau_cont.ps} \\

\end{tabular}
\caption{Negative lag frequency ($\nu_{\rm lag}$) vs $\mathit{M}_{\mathrm{BH}}$ (\emph{left panel}) and absolute amplitude ($\mid \tau\mid$) vs  $\mathit{M}_{\mathrm{BH}}$ (\emph{right panel}) trends (error bars represent the 1-$\sigma$ confidence interval). The lag frequencies and amplitudes are redshift-corrected. The best fit linear models (in log-log space) and the combined 1-$\sigma$ error on the slope and normalization are overplotted as continuous and dotted lines.
 The dashed lines in the right panel represent (from bottom to top) the light crossing time at 1$\mathit{r}_{\mathrm{g}}$, 2$\mathit{r}_{\mathrm{g}}$, and 6$\mathit{r}_{\mathrm{g}}$ as a function of mass.}
\label{fig:corr}

\end{figure*}
Thus, for each source we simulated 1000 pairs of correlated stochastic light curves, representing the hard and soft bands. We generated the light curves using the method of Timmer \& K\"{o}nig (1995), with different PSDs for each band but choosing the same random number seed  for each light curve in the pair so that the initial light curves are 100 per cent coherent.
In the simulations we assume the same source and background count rates and same Poisson noise level as the real data in the selected soft and hard energy bands.
 The intrinsic variability power of the source within each energy band have been estimated by fitting the corresponding PSDs (Fig. \ref{fig:psd}) with a simple power law plus constant (accounting for the Poisson noise contribution) model, and integrating over the sampled frequency range. 
The slope of the power law in the two bands has been fixed to the mean values obtained by Gonz\'{a}lez-Mart\'{i}n \& Vaughan (2012) using the same fitting model, i.e. $\alpha=$-2 for the soft band and $\alpha=$-1.7 for the hard band. The derived best fit parameters have been used to define the underlying PSDs of the simulated light curves. The latter are assumed to break to a slope of $\alpha=$-1 (e.g. Edelson \& Nandra 1999; Uttley et al 2002; Markowitz 2010) below a characteristic frequency.
The break frequency of each source has been estimated by rescaling the break frequency of 1H0707-495 (Zoghbi et al 2010) for the mass of the source, according to the scaling relationship provided by McHardy et al (2006), using the BH masses listed in Table \ref{tab:obs}, and discarding the dependence on the mass accretion rate\footnote{Assuming a linear scaling of the break frequency with the mass accretion rate (McHardy et al 2006), and using values tabulated in Ponti et al (2012) for the bolometric luminosity, this approximation introduces a negligible uncertainty, of a factor $\sim 2-3$, on the estimated break frequency. The latter uncertainty is even more negligible if the dependence of the break frequency on the mass accretion rate is weaker, as argued by Gonz\'{a}lez-Mart\'{i}n \& Vaughan (2012).}. 
For the sources in common, the estimated values are in agreement with results by Gonz\'{a}lez-Mart\'{i}n \& Vaughan (2012), and in most of the cases the PSD high frequency break lies below the analysed frequency range. It is worth noting that considering slightly different values for the PSD parameters (e.g. a steeper high frequency PSD slope, e.g. Vaughan et al 2011) does not change the results here.\\
Poisson noise contribution has been accounted for by adding Gaussians to each simulated light curves, with variance equal to the standard deviation of the mean count rate (the latter being estimated from the best fit constant level of Poisson noise in the PSDs).
A zero-phase lag was imposed on each pair of simulated light curves, meaning that every fluctuation in the resulting lag frequency spectrum above and below the zero-lag level is due to statistical noise. 
As previously mentioned, the effect of counting noise is to produce a deviation of the coherence function from 1, resulting in a drop to zero-value at high frequencies, where the Poisson noise variability power dominates. This effect is reproduced in the simulated data by adding the proper level of Poisson noise. However, the low-frequency modes can still be affected by an extra {\it intrinsic} fraction of incoherent signal (e.g. in 1H0707-495, Zoghbi et al 2010, and in REJ1034+396,  Zoghbi \& Fabian 2011, this drop in coherence has been attributed to the transition between two different variability processes). We accounted for it by adding a fraction $1-\gamma_{I}^2(\nu)$ of uncorrelated signal to every pair of simulated light curves, where $\gamma_{I}^2(\nu)$ represents the frequency-dependent fraction of intrinsic coherence as measured from the data.\\
The resulting simulated light curves were used to compute lag frequency spectra, adopting the same sampling (e.g. duration of each light curve segment, time resolution) and rebinning (e.g. logarithmic rebinning factor of the lag spectrum, minimum number of counts per bin) as in the real data. 
We then defined a sliding-frequency window, containing the same number, $N_{w}$, of consecutive frequency bins as those forming the observed negative lag profile in the data (i.e. typically $N_{w}$=3). 
The figure of merit $\chi=\sqrt{\sum (\tau/\sigma_{tau})^2}$ within the sliding window is computed at each step and the maximum value recorded for all the simulated lag frequency spectra. In this procedure the full-range of sampled frequencies is swept, with the exception of those where the measured coherence drops to zero as a consequence of Poisson noise (this cut-off being around $10^{-3}-10^{-2}$ Hz, see following section). Since our aim is to estimate the probability of recovering the observed lag profile only by chance, we registered the number of times $N_{w}$-consecutive negative lag points are observed in the simulated data with a figure of merit $\chi$ exceeding the real one. The estimated significances for the 15 detected soft lags are reported (in parentheses) in Table \ref{tab:obs}. In most of the sources we registered a mild decrease of the inferred significances with respect to those obtained with the standard statistical tests. Specifically, probabilities should be multiplied by a factor $\sim 0.97-0.99$, depending on the quality of the data. However, the corrected values are still consistent with lying above the adopted $2\sigma$ detection threshold.\\

\subsubsection{Poisson noise}
\label{sec_pois}
Counting noise gives an important contribution at high frequencies, where the intrinsic variability power of the source decreases to values comparable to the Poisson noise component. Poisson noise adds to the signal as an incoherent component, whose phase (i.e. the argument of the cross spectrum) is uniformly and randomly distributed in the range $[-\pi,+\pi]$. When combined with the intrinsic cross spectrum vector, this component increases the spread in the phase. Fourier phase lags are limited to the range of values $[-\pi,+\pi]$, corresponding to the condition $|\tau|\leq 1/2\nu$ on the time lag amplitude. The latter limits on $\tau$ have been marked in Fig. \ref{fig:lags} as continuous red curves. This implies that, if the lag values within each frequency bin are highly spread (e.g. as a consequence of low variability power and/or low intrinsic coherence, and so higher Poisson noise fraction) they tend to distribute uniformly within these limits, since the tails of the distribution are cut at $\tau= \pm 1/2\nu$ and reflected within the permitted range of values. In this regime, the uncertainty on the lag is of the order of the range of permitted values, thus precluding the detection of any intrinsic lag. 
However, rebinning the cross spectrum over contiguous frequencies and/or over different light curve segments, reduces this effect, since the Poisson noise components to the cross spectrum vectors tend to cancel out in the averaging process. In other words this allows the detection of time lags up to frequencies higher than the frequency at which the variability power of the source is of the order of the Poisson noise component. 
Hence, the frequency $\nu_{\rm Poiss}$  at which Poisson noise starts dominating the lag-frequency spectra is given by the frequency at which the spread in the measured time lag (defined by the 1$\sigma$ contour plots in Fig. \ref{fig:lags}) is of the order of the standard deviation (dotted red curves in Fig. \ref{fig:lags}) of a uniform distribution defined on the interval $\tau=[-1/2\nu,+1/2\nu]$. As is clear from Fig. \ref{fig:lags}, the data are well-sampled enough that Poisson noise and intrinsic coherence do not lead to errors approaching the $\pm \pi$ limits of the phase lag.\\
To determine $\nu_{\rm Poiss}$  following a more rigorous approach we relied on Monte Carlo simulations. 
With respect to previous simulations we adopted the different approach of studying the effects of Poisson noise on the cross spectrum terms at each frequency. To this aim we generated 1000 realisations of the Poisson noise cross spectrum vector. The randomization of the corresponding phase and amplitude is obtained by drawing two Gaussian distributed random numbers and using them as the real and imaginary part. The standard deviation of the two Gaussian distributions is a function of the measured coherence, $\gamma^2$, and the number of samples into each frequency bin ($N=k*m$, where $k$ is the number of frequencies and $m$ is the number of light curve segments averaged over) and is given by $\delta=\sqrt{(1-\gamma^2)(2N\gamma^2)^{-1}}$. 
Each simulated vector is then added to the intrinsic cross spectrum vector. 
We applied this procedure to all the detected soft lags, assuming a constant soft/negative time lag over all the sampled frequency range, whose amplitude is fixed to the minimum value in the observed lag profile, and checked at which frequency, $\nu_{\rm Poiss}$, the distribution of phases starts to tend to a uniform distribution and the average phase goes to zero. In most of the sources the estimated $\nu_{\rm Poiss}$ is $\gsim 0.01$Hz, falling well above the highest frequency sampled. Only in four cases (i.e. Mrk 335, NGC 3516, NGC 5548, and Mrk 841) $\nu_{\rm Poiss}$ is lower ($> 10^{-3}$ Hz). 
Thus the lag minimum is always detected significantly (a factor $\sim 3-20$) below $\nu_{\rm Poiss}$, in agreement with results of zero-lag Monte Carlo simulations, whereby the observed negative lags are not due to spurious noise features.\\

\begin{figure} 
\centering

\begin{tabular}{p{8cm}}
\includegraphics[height=3.5cm,angle=270]{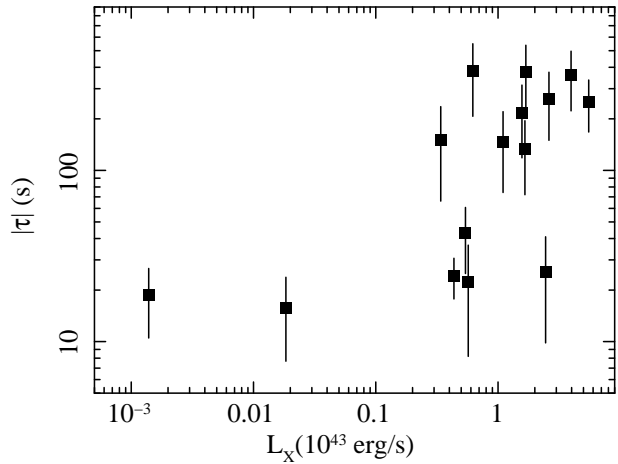} \\

\end{tabular}
\caption{Absolute amplitude of the negative lags ($\mid \tau\mid$) vs  2-10 keV luminosity (errors bars represent the 1-$\sigma$ confidence interval).}
\label{fig:Lx}
\end{figure}

\section{Lag correlations}
\label{lagmass}

The measured lag-frequency spectra have similar profiles, with the negative lag shifting to lower frequencies as the black hole mass of the source increases (see Fig. \ref{fig:lags}, where the plots are in order of increasing mass of the source). It is thus interesting to test the hypothesis 
 of a $\mathit{M}_{\mathrm{BH}}$ dependence of the soft lag characteristic time-scales (i.e. amplitude and frequency).\\
The sources showing a soft/negative lag have widely different mass values, spanning about 2.5 orders of magnitude in estimated $\mathit{M}_{\mathrm{BH}}$.
The $\mathit{M}_{\mathrm{BH}}$ values adopted in this paper are taken from the literature (see Table \ref{tab:obs} for references) and are preferably those obtained from reverberation mapping and stellar velocity dispersion techniques.
In all the other cases (5 out of 15) we considered estimates obtained from the empirical relation between the optical luminosity at 5100\AA\ and the broad line region size ($\mathit{R}_{BLR}$, e.g. Bentz et al 2009). 
 The unknown mass uncertainties have been replaced with the estimated dispersion of the adopted relation for mass determination (i.e. 0.5 dex in the case of single epoch methods, and 0.4 dex for the others).\\
Fig. \ref{fig:corr} shows the lag frequency and amplitude vs mass ($\nu_{\rm lag}$ vs $\mathit{M}_{\mathrm{BH}}$ and $\tau$ vs $\mathit{M}_{\mathrm{BH}}$) on a logarithmic scale. In the plots the $\nu_{\rm lag}$ and $\tau$ values (and their uncertainties) correspond to the frequency and amplitude of the minimum, single-point in the negative lag profile (the effects of statistical fluctuations in the position of the lag minimum are further discussed in Sect. \ref{sec_corr_test}), once corrected for the redshift of the source.
The fit of these data sets with a constant is unsatisfactory, yielding $\chi^2=213$ and $\chi^2=41$ (14 degrees of freedom) respectively for the lag frequency and amplitude.
To statistically assess whether the pairs of parameters are correlated we computed the Spearman's rank correlation 
coefficient, $\rho$. The test yielded $\rho\sim -0.79$ and $\rho \sim 0.90$ respectively for the $\nu_{\rm lag}$ 
and $\tau$ parameters, both corresponding to a correlation with significance $\gsim 4\sigma$ (null-hypothesis probabilities of $\sim 4\times10^{-4}$ and $\sim 5\times10^{-6}$, respectively), with the correlation between $\tau$ and BH mass being the most significant.\\
It is worth noting that the lag profile in the large mass sources is not well-sampled, since the data are limited (at the lowest frequencies) by the duration of the observations. Hence, we checked whether the correlation still holds when considering these points as upper/lower limits for the lag frequency/absolute amplitude, by computing generalised Kendall's rank correlation statistics for censored data sets (e.g. Isobe et al 1986). Although slightly lower, the significance of the correlation is still relatively high ($>3\sigma$).\\
Following Bianchi et al (2009b), we applied a least squares linear regression approach to derive constraints on the functional dependence of $\nu_{\rm lag}$ and $\tau$ from $\mathit{M}_{\mathrm{BH}}$. To minimize the uncertainty in the best fit model we carried out 1000
Monte Carlo simulations, whereby the x- and y-axis coordinates of each experimental point were substituted by two random values drawn from two Gaussian distributions with means equal to the coordinates of the data point on each axis and the associated statistical uncertainty as standard deviation (see Bianchi et al 2009b).
 Each simulated data set was fitted in log-log space with a linear model. The mean of the slopes and intercepts were used to define our 
best fit model, while the mean standard deviation represents the uncertainty of the best fit parameters. 
The results of the fits are $\mbox{Log }\nu_{\rm lag}=-3.50[\pm 0.07]-0.47[\pm 0.09]\mbox{ Log}(\mathit{M}_{7})$ and $\mbox{Log }\mid\tau \mid=1.98[\pm 0.08]+0.59[\pm 0.11]\mbox{ Log}(\mathit{M}_{7})$, where $\mathit{M}_{7}=\mathit{M}_{\mathrm{BH}}/10^{7} \mathit{M}_{\odot}$. The estimated scatter around the best fit model is $\sigma_{s}\sim 0.19$ and $0.23$, respectively for the $\tau$--$\mathit{M}_{\mathrm{BH}}$ and $\nu_{\rm lag}$--$\mathit{M}_{\mathrm{BH}}$ relations. The fit significantly improves with respect to the fit with a constant model, yielding a $\chi^2\sim 66$ and $\chi^2\sim 17$ (respectively for the lag frequency and amplitude data sets) with the addition of one parameter (corresponding to a $\gsim$99.9 per cent F-test probability).
 It is worth noting that the small scatter in the relation is consistent with being mostly due to the uncertainty in BH mass determination, being of the same order of the intrinsic scatter in reverberation mass estimates.\\
The best fit and the 1-$\sigma$ combined uncertainty on the model parameters have been overplotted on the data in Fig. \ref{fig:corr}. We note that, although statistically not consistent with a linear relation, the best fit slopes increase and become consistent with linear scaling, if the data from the larger mass objects are treated as upper/lower limits for the lag frequency/absolute amplitude. Moreover, as pointed out in Sect. \ref{sec:lag_specs}, the simulations of the observed lag profile revealed that red noise leakage plays a role in decreasing the intrinsic amplitude of the lag at low frequencies, thus affecting mostly the soft lag in high mass sources. A rough estimate yields a decreasing factor of $\sim$2, with fluctuations depending on the shape of the PSD. Indeed, this effect is less important in low mass sources, where the amount of power below the lowest sampled frequency is smaller. Overall, correcting the observed $\tau$--$\mathit{M}_{\mathrm{BH}}$ correlation for this bias would result in a steeper (i.e. with best fit value of $0.72\pm0.11$) function of BH mass.\\
Finally, we checked whether the lag time-scales exhibit some dependence on the source X-ray luminosity
between $\mathit{E}=2-10$ keV (see Fig. \ref{fig:Lx}), but a less significant correlation was found ($<3 \sigma$ and $<2\sigma$ respectively for $\tau$ and $\nu_{\rm lag}$), which further decreases ($<2\sigma$) when looking at the correlation with the bolometric luminosity (as computed using the correction
factor by Marconi et al 2004).
These results support the hypothesis that the main parameter driving the correlation is the BH mass.\\

\begin{figure*}
\centering
\begin{tabular}{p{8.cm}p{8.cm}}
\includegraphics[height=8.2cm,angle=90]{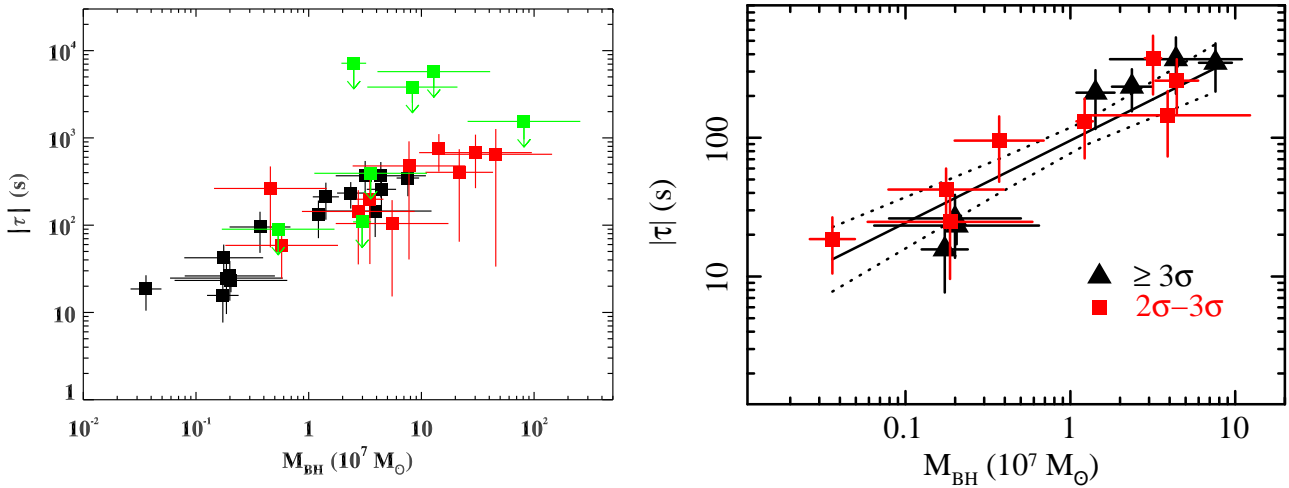} & \includegraphics[height=7.2cm,angle=270]{figure/corr_Mtau_cont23sigma.ps} \\
\end{tabular}
\caption{\emph{Left panel:} The 10 marginal soft lags with significance between 1-2$\sigma$ (red data points) and the 7 with significance $<1\sigma$ (green upper limits) overplotted to the 15 detections at $>2\sigma$ (black data points). \emph{Right panel:} $\mid \tau\mid$ vs  $\mathit{M}_{\mathrm{BH}}$ separate trends for high significance (black triangles) and low significance detections (red squares). Overplotted is the best fit linear model in log-log scale for the whole data set.}
\label{taucorr23sigma}

\end{figure*}

{\bf {\it Non-detections:}} we checked whether the absence of a soft/negative lag in the remaining 17 sources may be real or due to poor statistics. Hence, for each lag-frequency spectrum we recorded the most negative lag amplitude and its frequency in the interval where the coherence is 
not consistent with zero. We collected a total of 10 marginal soft lags with significance between 1-2$\sigma$ and 7 with significance $<1\sigma$ (examples are shown in Fig. \ref{fig:lags2}, while details on the 17 sources are listed in Table \ref{tab:obs2}). Results are shown in left panel of Fig. \ref{taucorr23sigma} (where, for brevity, only the $\tau - \mathit{M}_{\mathrm{BH}}$ data points are displayed, overplotted on the 15 soft lag detections at $>2\sigma$ confidence level) where the $>1\sigma$ detections are plotted with error bars, while the $<1 \sigma$ are marked as 90 per cent upper limits. The agreement with the overall correlation is clear, with the significance increasing to $>5 \sigma$ confidence level ($\rho \sim 0.73$ and $0.86$, respectively for the frequency and amplitude vs BH mass correlations). 
We conclude that none of the remaining 17 sources is, with the available data, significantly outside the observed correlation and that all the sources are consistent with showing the same lag properties. The low detection significance of the soft lag in these sources is mainly due to poor statistics issues (e.g. limited exposure, small number of observations, low variability power, etc.), which are most relevant in large mass sources, where the expected lag frequency is of the order of the minimum sampled frequency.

\subsection{Tests of the correlation}
\label{sec_corr_test}
The correlations between the soft lag and BH mass presented in this paper have been derived assuming the minimum, single-point in the negative lag profile as representative of the intrinsic soft lag time-scale. This choice is arbitrary, thus, to be more conservative, we estimated the effects of statistical fluctuations in the position of the soft lag minimum on the observed $\nu_{\rm lag}$-$\mathit{M}_{\mathrm{BH}}$ relation. To this aim we made use of the Monte Carlo simulations generated to determine the confidence contours of the lag profile (see Sect. \ref{sec:lag_specs}). These simulations assume as a template an interpolated lag function that describes the measured lag, whereby any deviation of the minimum from the observed one is due to statistical fluctuations. Hence, for each simulated lag profile we recorded the frequency of the minimum time lag, and used them to infer a mean frequency and standard deviation. These values were correlated with BH mass, to check if the observed correlation still holds.  We found that statistical fluctuations in the position of the lag minimum have almost negligible effects, only slightly reducing the Spearman's rank correlation coefficient to $\rho \sim 0.78$ (i.e. bringing the null-hypothesis probability from $\sim 4\times 10^{-4}$ to $\sim 7\times 10^{-4}$), and increasing the scatter around the best fit model from $\sim$0.23 to $\sim$0.26. The fact that the difference with respect to results obtained using $\nu_{\rm lag}$ measured values is small is due to the fact that, in general, the soft lag profiles are not very broad in frequency. Thus the effects of statistical fluctuations on the position of the minimum within the soft lag range of frequencies are negligible as compared to the scatter in the relation introduced by the uncertainties in the BH mass, and does not affect the stronger correlation between $\tau$ and BH mass.\\
It is worth noting that a good correlation is still present even when only the highest significance lags are taken into account. Indeed, carrying out a separate Spearman test on the $\tau$ vs $\mathit{M}_{\mathrm{BH}}$ values for the $\gsim 3\sigma$ detections, and for those between $2-3\sigma$, we obtain a $\rho \sim 0.9$ in each case, which corresponds to a correlation significance of $\sim 99.8$ per cent. This is clearly visible in Fig. \ref{taucorr23sigma} (right panel) where sources belonging to the two different groups are displayed using different colours/symbols. Moreover, if only the $\gsim 3\sigma$ detections lag points are taken into account in the fits with a linear model (in log-log space), the best fit slope parameter (i.e. $0.75\pm0.18$) is steeper than the value obtained from the whole data set (Sect. \ref{lagmass}). Correction for red noise leakage leads to a further steepening of the correlation (with best fit value of $0.95\pm0.19$).\\

\section{Discussion}

The systematic analysis of X-ray lags presented in this paper significantly increases the number of soft lags so far detected in AGN (15 out of 32 sources). The main result is the discovery of a highly significant ($\gsim 4\sigma$) correlation between the time-scales (frequency and amplitude) of the detected soft/negative lag and the BH mass.
Moreover, data from the remaining 17 sources are consistent with the observed correlation, the significant detection of a soft lag being precluded only by poor statistics in these observations. Hence, we conclude that a soft lag, scaling with the mass of the BH, cannot be excluded in any of the 32 sources of our sample.
 This result confirms our previous inference (De Marco et al 2011), whereby the negative lag detected in PG 1211+143 was suggested to represent the large mass counterpart of negative lags detected in lower mass sources (such as 1H0707-495). Most importantly it highlights a fundamental property of soft lags, namely the fact that 
they depend on the BH mass. For example, in a reverberation scenario, this is naturally expected, given that the gravitational radius light crossing time ($t_c=r_g/c$, where $r_g=GM/c^2$ is the gravitational radius) scales linearly with the mass of the central object.
 Although the inferred best fit slopes for the observed lag-$\mathit{M}_{\mathrm{BH}}$ trends are statistically different from a linear relation, we notice that the data are biased against an intrinsically steeper dependence. Indeed, a steeper correlation is observed if the bias due to red noise leakage is taken into account (Sect. \ref{sec:lag_specs}). The same correction leads to a one-to-one BH mass vs lag correlation if only the highest significance ($\geq 3\sigma$) detections are considered in the fit (Sect. \ref{sec_corr_test}).
Moreover, the large mass sources are sensitive to the length of the observation, which does not allow us to measure the lag profile over frequencies lower than the inverse of the observation length.
 Albeit not introducing any spurious correlation, the effect of these biases is to induce a flattening of the intrinsic one. Indeed, when the large BH mass data points are treated as upper/lower limits the slope of the correlations becomes even steeper (Sect. \ref{lagmass}).
A sample including longer and more sensitive observations would be necessary to accurately determine the real slope of the relation.\\
A thorough study of the implications of the lag-$\mathit{M}_{\mathrm{BH}}$ correlation, as inferred from our analysis, on the proposed soft lags models is beyond the aim of this paper. 
However, it is worth noting that a BH mass dependence of the characteristic time-scales is naturally expected in standard accretion disc models. 
 In Fig. \ref{fig:corr} we overplotted as a reference the light crossing time at 1$\mathit{r}_{\mathrm{g}}$, 2$\mathit{r}_{\mathrm{g}}$, and 6$\mathit{r}_{\mathrm{g}}$ as a function of BH mass. The observed soft lags roughly lie in this range of time-scales, meaning that the involved distances are quite small and, again, mass-dependent. 
Moreover, the evidence for a soft lag being present in such a large number of sources is at odds with expectations from models that need the single objects to be on a special line of sight for the measured lag to be observable (e.g. Miller et al 2010).
It is worth noting that, although the observed amplitudes of the soft lags roughly agree with the light-crossing time at small radii, associating a measured lag with a particular radius is not trivial (Wilkins \& Fabian 2013). For instance, the observed lag is likely to be smaller than the intrinsic one due to the cross-contamination of primary and delayed spectral components in the two bands of interest (e.g. Miller et al 2010), which introduces a dilution factor for the lag (see e.g. Kara et al 2013a, 2013b). Moreover, relativistic effects (e.g. Shapiro delay) are likely to play a major role which prevents us from trivially matching an observed time-delay with a given distance. Another important aspect has been recently pointed out by Kara et al (2013b) who show that the frequency and amplitude of the soft lag in IRAS 13224-3809 are in fact flux--dependent. If this behaviour is ubiquitous, it will affect the measured lags at least in cases where the X-ray observation is not long enough to sample appropriately all different X--ray flux levels of a given source. This is likely to mostly affect the measured lags for the slowest varying highest BH mass sources in which a particular X-ray observation may be biased towards a particular flux state.\\
Overall these results are consistent with a scenario whereby the delayed soft excess emission originates in the innermost regions of the accretion disc, and is triggered by a compact central source of hard X-rays.
Hence, a thorough understanding of soft lags properties will allow us to probe the physics and geometry of the innermost regions of AGN.

\section*{Acknowledgments}

This work is based on observations obtained with XMM-{\it Newton}, an ESA science mission with instruments and contributions directly funded by ESA Member States and NASA. BDM, MC, and MD thank financial support from the ASI/INAF contract I/009/10/0.
BDM and GM thank the Spanish Ministry of Science and Innovation (now Ministerio de Econom\'{i}a y Competitividad) for financial support through grant AYA2010-21490-C02-02. GP acknowledges support via an EU Marie Curie Intra-European Fellowship under contract no. FP7-PEOPLE-2009-IEF-254279. BDM acknowledges A. Lovato for useful discussions.
The authors thank the anonymous referee for valuable comments which contributed to significantly improve the paper.

\begin{table*}
\caption{Sources with a soft/negative lag detection at $>2\sigma$ confidence level. (1) Name; (2) XMM-{\it Newton} observations; (3) Nominal and effective exposure; (4) Logarithm of the estimated black hole mass, uncertainty, mass estimate technique (R: reverberation mapping, V: stellar velocity dispersion, E: empirical $\mathit{L}_{5100\AA}$ vs $\mathit{R}_{BLR}$ relation), and references ([a] Nelson \& Whittle 1995; [b] Oliva et al 1999; [c] Kaspi et al 2000; [d] Vestergaard 2002; [e] Bian \& Zhao 2003; [f, g] Peterson et al 2004, 2005; [h] Zhang \& Wang 2006; [i] Wang \& Zhang 2007; [l] Bentz et al 2009; [m] Zhou et al 2010; [n] Denney et al 2010); (5) Soft and hard X-ray energy bands used for the lag computation; (6) soft/negative lag significance as obtained from the combination of the consecutive points forming the soft lag profile, and, within parenthesis, from Monte Carlo simulations.}
\label{tab:obs}
\centering
\vspace{0.2cm}
\begin{scriptsize}
\begin{tabular}{c c c c c c}
\hline          

 Name            &        Obs ID     &  Exposure [effective] (ks)  & Log ($\mathit{M}_{\mathrm{BH}}$) & $\Delta \mathit{E}_{\mathrm{soft/hard}}$ (keV)  & Significance \\
   (1)    &   (2)  &  (3)  &  (4)  &   (5)  &   (6)  \\ 
\hline

NGC 4395         &      0142830101   &    113  [90] &  5.56$\pm$0.14 R, g & 0.3-1/1-5 & 99.3\% (98.0\%) \\

NGC 4051  &   0109141401    &  122  [100] & 6.24$\pm$0.14 R, n & 0.3-1/1-5 & 99.9\% (99.5\%)\\
                 &     0157560101    &  52 [42]& &  & \\                 
                 &     0606320101    &  46  [45]  & &  & \\	
                 &     0606320201    &  46  [42] & &  & \\
                 &     0606320301    &  46  [21] & &  & \\	
                 &     0606320401    &  45  [18] & &  & \\	
                 &     0606321301    &  33  [39] & &  & \\
                 &     0606321401    &  42  [40] & &  & \\	
                 &     0606321501    &  42  [34] & &  & \\	
                 &     0606321601    &  42  [41] & &  & \\	
                 &     0606321701    &  45  [38] & &  & \\
                 &     0606321801    &  44  [40] & &  & \\
                 &     0606321901    &  45  [36] & &  & \\
                 &     0606322001    &  40  [37] & &  & \\
                 &     0606322101    &  44  [24] & &  & \\
                 &     0606322201    &  44  [41] & &  & \\
                 &     0606322301    &  43  [42] & &  & \\

Mrk 766          &     0109141301    & 130 [105] &  6.25$\pm$0.35 R, l & 0.3-0.7/1.5-4 & 99.6\% (97.6\%)\\
                 &     0304030301    & 99   [98] & &  & \\
                 &     0304030401    & 99   [93] & &  & \\ 
                 &     0304030501    & 96   [93] & &  & \\
                 &     0304030601    & 99   [85] & &  & \\
                 &     0304030701    & 35   [16] & &  & \\

Ark 564          &     0006810101    & 34 [11] & 6.27$\pm$0.50 E, h & 0.3-1/2-5 & 97.0\% (98.4\%)\\
                 &     0006810301    & 16  [12] & &  & \\
                 &     0206400101    & 102 [99] & &  & \\

MCG-6-30-15       &     0111570101    & 47  [10] & 6.30$\pm$0.40 V/E, b, m & 0.3-0.9/1.5-3 & 99.9\% (99.5\%)\\ 
                 &     0111570201    &  66   [51]   & &  & \\             
                 &     0029740101    &  89  [80] & &  & \\
                 &     0029740701    &  129 [122] & & \\
                 &     0029740801    &  130 [124] & & \\

1H 0707-495      &     0511580101    & 124 [122] & 6.31$\pm$0.50 E, e & 0.3-1/1-4 & $>$99.9\% ($>$99.9\%)\\  
                 &     0511580201    & 124  [97] & &  & \\
                 &     0511580301    & 123  [100] & &  & \\
                 &     0511580401    & 122   [86] & &  & \\

RE J1034+396     &  0506440101	 &  93 [84] & 6.57$\pm$0.27 E, m & 0.3-1/1.5-4.1 & 98.4\% (96.0\%) \\
                 &  0561580201   &  70  [51] & &  & \\
                 &  0655310101   &  52 [20] & &  & \\
                 &  0655310201   &  54 [32] & &  & \\

NGC 7469     &   0112170301  & 25 [18] &  7.09$\pm$0.05 R, f & 0.3-1.5/1.5-5 & 97.0\% (97.0\%)\\
             &   0112170101  & 19  [23] & &  & \\
             &   0207090201  & 79  [84] & &  & \\
             &   0207090101  & 85  [78] & &  & \\

Mrk 335      &   0306870101  & 133 [118] & 7.15$\pm$0.12 R, f & 0.3-0.6/3-5 & 99.7\% (98.6\%) \\ 
             &   0600540501  & 83  [110] & &  & \\
             &   0600540601  & 132 [81] & &  & \\

PG 1211+143  &   0112610101 & 56 [53] &  7.37$\pm$0.12 R, c & 0.3-0.7/1.5-5 & 99.9\% (99.8\%)\\ 
             &   0208020101 &  60  [34] & &  & \\
             &   0502050101 & 65   [46] & &  & \\
             &   0502050201 & 51  [20] & &  & \\

NGC 3516     & 0107460601 & 124 [75] & 7.50$\pm$0.05 R, n & 0.3-1/1-5 & 99.3\% (97.2\%)\\ 
             & 0107460701 & 130 [116] & &  & \\
             & 0401210401 & 52  [52] & &  & \\
             & 0401210501 & 69  [60] & &  & \\
             & 0401210601 & 69  [60] & &  & \\
             & 0401211001 & 69  [42] & &  & \\

NGC 6860     &   0552170301 & 123 [117] & 7.59$\pm$0.50 E, i & 0.3-1/1-5 & 98.8\% (97.8\%)\\ 

Mrk 1040     &   0554990101 & 91 [89] & 7.64$\pm$0.40 V, a, m & 0.3-1/1-4 & 99.7\% (98.6\%) \\ 

NGC 5548     &   0089960301 & 96 [84] & 7.65$\pm$0.13 R, n & 0.3-0.9/1.5-4.5 & 99.1\% (97.7\%)\\ 

Mrk 841      &   0070740101 & 12 [11] & 7.88$\pm$0.10 E, d & 0.3-1/1-5 &  $>$99.9\% (99.9\%)\\ 
             &   0070740301 & 15 [12] & &  & \\
             &   0205340201 & 73 [42] & &  & \\
             &   0205340401 & 30 [15] & &  & \\
\hline  
\end{tabular}

\end{scriptsize}

\end{table*}

\begin{table*}
\caption{Sources with a marginal negative lag detected at $<2\sigma$ confidence level. (1) Name; (2) XMM-{\it Newton} observations; (3) Nominal and effective exposure; (4) Logarithm of the estimated black hole mass, uncertainty, mass estimate technique (R: reverberation mapping, V: stellar velocity dispersion, E: empirical $\mathit{L}_{5100\AA}$ vs $\mathit{R}_{BLR}$ relation), and references ([a] Oliva et al 1995; [b] Peterson et al 2004; [c] Wang, Watari \& Mineshige 2004; [d] Zhang \& Wang 2006; [e] Zhou \& Wang 2005; [f] Vestergaard \& Peterson 2006; [g] Jin et al 2009;  [h] Markowitz 2009; [i] Wang, Mao \& Wei 2009; [l] Winter et al 2009); (5) Soft and hard X-ray energy bands used for the lag computation; (6) soft/negative lag significance as obtained from the combination of the consecutive points forming the soft lag profile.}
\label{tab:obs2}
\centering
\vspace{0.2cm}
\begin{scriptsize}
\begin{tabular}{c c c c c c}
\hline          

 Name            &        Obs ID     &  Exposure [effective] (ks)  & Log ($\mathit{M}_{\mathrm{BH}}$) & $\Delta \mathit{E}_{\mathrm{soft/hard}}$ (keV)  & Significance \\
   (1)    &   (2)  &  (3)  &  (4)  &   (5)  &   (6)  \\ 
\hline

 IRAS F12397+3333 & 0202180201 & 80 [65] &  6.66$\pm$0.50 E, d & 0.3-1/1-5 & 90\% \\
 IRAS 13224-3809  & 0110890101 & 64 [57] &  6.76$\pm$0.50 E, e & 0.3-1/1-4 & 93\% \\
 Mrk 1502         & 0110890301 & 22 [20] &  7.44$\pm$0.50 E, f & 0.3-1/1-5 & 84\% \\ 
                  & 0300470101 & 86 [74] &       & & \\
 Mrk 279          & 0083960101 & 33 [18] &  7.54$\pm$0.12 R, b & 0.3-1/1-5 & 77\%\\
                  & 0302480401 & 60 [40] &      & & \\
                  & 0302480501 & 60 [48] &      & & \\
                  & 0302480601 & 38 [27] &      & & \\
 IRAS 13349+2438  & 0096010101 & 65 [42] &  7.74$\pm$0.50 E, c & 0.3-0.8/0.8-2& 90\%\\ 
                  & 0402080201 & 48  [22] & & & \\
                  & 0402080301 & 69 [60] & & & \\
 RX J0136.9-3510  & 0303340101 & 54 [43] &  7.89$\pm$0.50 E, g & 0.3-1/1-5 & 86\%\\ 
 Mrk 509          & 0130720101 & 32 [30] &  8.16$\pm$0.04 R, b & 0.3-1/1-4 & 89\%\\
                  & 0130720201 & 44 [40] &  & & \\
                  & 0306090201 & 86 [85] &  & & \\
                  & 0306090301 & 47 [46] &  & & \\
                  & 0306090401 & 70 [69] &  & & \\
 IC 4329a         & 0147440101 & 136 [120] &  8.34$\pm$0.30 V, h & 0.3-0.8/1-4 & 77\%\\ 
 ESO 198-G24      & 0067190101 & 34 [30] &  8.48$\pm$0.50 E, e & 0.3-1/1-5 & 93\% \\
                  & 0305370101 & 122 [112] &  & & \\
 ESO 511-G030     & 0502090201 & 112 [105] &  8.66$\pm$0.50 E, l & 0.3-1/1-5 & 71\%\\ 
 NGC 4593         & 0059830101 & 87 [74] &  6.73$\pm$0.43 R, b & 0.3-1/1-5 & $<$68\%\\
 Mrk 110          & 0201130501 & 47 [47] &  7.40$\pm$0.11 R, b & 0.3-1/1-5 & $<$68\%\\
 NGC 3783         & 0112210101 & 40 [37] &  7.47$\pm$0.08 R, b & 0.3-1/1-5 &$<$68\%\\ 
                  & 0112210201 & 138 [125] & & & \\
                  & 0112210501 & 138 [124] & & & \\
 IRAS 05078+1626  & 0502090501 & 62 [56] &  7.55$\pm$0.50 E, i & 0.3-1/1-5 & $<$68\%\\
 MCG -5-23-16     & 0112830401 & 25 [22] &  7.92$\pm$0.40 V, a & 0.3-1/1-5 & $<$68\%\\
                  & 0302850201 & 132 [102] & & & \\
 Mrk 704          & 0300240101 & 22 [21] &  8.11$\pm$0.50  E, i & 0.3-1/1-5 & $<$68\%\\ 
                  & 0502091601 & 98 [86] & & & \\
 PDS 456          & 0041160101 & 47 [41] &  8.91$\pm$0.50 E, e & 0.3-1/1-5 & $<$68\%\\
                  & 0501580101 & 92 [89] & & & \\
                  & 0501580201 & 90 [86] & & & \\
\hline  
\end{tabular}

\end{scriptsize}
\end{table*}

\end{document}